\documentstyle[twocolumn,aps,epsf]{revtex}           
\begin{document}                                     
\draft                                               
\twocolumn[\hsize\textwidth\columnwidth\hsize\csname 
@twocolumnfalse\endcsname

\title{Adaptive Mesh Refinement Computation of Solidification Microstructures 
using Dynamic Data Structures} \author{Nikolas Provatas$^{1,2}$, Nigel
Goldenfeld$^{1}$, and Jonathan Dantzig$^{2}$ }

\address{ $^1$University of Illinois at Urbana-Champaign,
Department of Physics \\1110 West Green Street, Urbana, IL,
61801 }

\address{ $^2$  University of Illinois at Urbana-Champaign,
Department of Mechanical and Industrial Engineering \\1206 West
Green Street, Urbana, IL, 61801 }

\date{\today} 
\vspace{2.0cm}

\maketitle
]

\noindent \underline{\hspace{8.5cm} }
\begin{abstract} 

We study the evolution of solidification microstructures
using a phase-field model computed on an adaptive, finite element  
grid.  We discuss the details of our algorithm and show that 
it greatly reduces the computational cost of solving the phase-field
model at low undercooling. In particular we show that the computational complexity 
of solving any phase-boundary problem scales with the interface arclength 
when using an adapting mesh. Moreover, the use of dynamic data 
structures allows us to simulate system sizes corresponding to experimental
conditions, which would otherwise require lattices greater that 
$2^{17}\times 2^{17}$ elements. We examine 
the convergence properties of our algorithm. We also present 
two dimensional, time-dependent calculations of dendritic evolution,
with and without surface tension anisotropy.  We benchmark our results 
for dendritic growth with microscopic solvability theory, finding them to 
be in good agreement with theory for high undercoolings. At low undercooling, 
however, we obtain higher values of velocity than solvability theory at 
low undercooling, where transients dominate, in accord with a heuristic 
criterion which we derive.

\end{abstract}
\noindent \underline{\hspace{8.5cm} }

\section{Introduction}

Modeling solidification microstructures has become an area of
intense study in recent years. The properties of large scale
cast products, ranging from automobile engine blocks to  aircraft 
components and other industrial applications, are strongly dependent on the physics
that occur at the mesoscopic and microscopic length scales
during solidification.  The main ingredient of the
solidification microstructure is the dendrite, a snowflake-like pattern of
solid around which solidification proceeds.   The microscopic 
properties of such cast products are determined by the length scales
of these dendrites, and for this reason understanding the mechanisms
for pattern selection in dendritic growth has attracted a great deal of 
interest from the experimental and theoretical community. 
In particular,  a great deal of research has
been undertaken to understand such issues as dendrite
morphology, shape and speed. Experiments on dendrite evolution
by Glicksman and coworkers on succinonitrile (SCN)
\cite{Hua81,Gli84}, and more recently Pivalic Acid (PVM)
\cite{Mic98}, as well as other transparent analogues of metals
have provided  tests of theories of dendritic growth, and have
stimulated considerable theoretical
progress\cite{Lan80,LanII80,Kes88}.  These experiments have
clearly demonstrated that in certain parameter ranges the
physics of the dendrite tip can be characterized by a
steady value for the dendrite tip velocity, radius of curvature and shape.  Away from
the tip the time-dependent dendrite exhibits the characteristic
sidebranching as it propagates.

The earliest theories of dendritic growth solved for the
diffusion field around a self-similar body of revolution
propagating at constant speed \cite{Iva47,Hor61}. In these
studies the diffusion field is found to determine the product of the
dendrite velocity and tip radius, but neither quantity by
itself. Adding capillarity effects to the theory predicts a
unique maximum growth speed \cite{Tem60}. Experiments, however,
have  shown that these theories do not represent the correct
operating state for real dendrites.

The introduction of local models of solidification brought
further insight to the steady state dendrite problem
\cite{Bro83,Ben83,Ben84,KessI84}. These models describe the
evolution of the interface, incorporating the physics of the
bulk phases into the governing equation of motion of the interface. A
remarkable breakthrough of these models was to show that a
steady-state dendrite velocity is obtained {\it only} if a source of
anisotropy (e.g., in the interfacial energy) is present during dendritic 
evolution.  The dendrite steady-state tip
velocities appear in  a discrete rather than continuous spectrum 
of values, making the role of anisotropy of great
importance in the description of the dendrite problem, both in
the local models and the full moving boundary problem
\cite{Kes88,Bre91,Pom91}.  It was further shown that only the
fastest of a spectrum of steady state velocities is stable, thus
forming the operating state of the dendrite.  This body of
theoretical work is generally known as {\it microscopic
solvability theory}.

Dendritic sidebranching is widely believed to be caused by
thermal fluctuations, which enter solidification models in the
form of random noise possessing specific features
\cite{Lan87,Kar93,Bre95}.  However, the precise mechanism of
sidebranch amplification does not seem to be fully understood.  It would
appear that the manner in which thermal noise is amplified may
depend on the overall dendrite morphology. For instance, it was
shown that noisy fluctuations traveling down a paraboloid of
revolution do not produce sidebrach amplitudes consistent with
experiments \cite{Lan87}, while fluctuations traveling down an
initially missile-shaped dendrite amplify into sidebranches
comparable to some experiments \cite{Bre95}. Karma also
investigated the addition of interface fluctuations
\cite{Kar93}. However, this source of noise  only becomes relevant
at high velocities.

The theoretical foundation around which most theories of
solidification are based is the time-dependent Stefan problem.
This theory describes the evolution of the thermal or solutal
diffusion field around the solidification front, along with two
accompanying boundary conditions. The first boundary condition
relates the velocity of the moving front to the difference in
thermal fluxes across the solid-liquid interface. The second,
called the Gibbs-Thomson condition, relates the
interfacial temperature to the the thermodynamic equilibrium 
temperature, the local interfacial curvature and
interface kinetics. The interface kinetics term adds a
non-equilibrium correction to the interface temperature, usually
assumed to be in local equilibrium for a given curvature.
Solving the Stefan problem numerically has traditionally
involved front tracking and lattice deformation to contain the
interface at predefined locations on the grid
\cite{Alm91,Try96}.  This method is generally complicated to
implement accurately and requires much effort. Moreover, it can
be inefficient in handling coalescence of two or more
interfaces.

The solution of the Stefan problem has been made more tractable
with the introduction of the {\it phase-field} model.  The 
phase-field model avoids this problem of front tracking by
introducing an auxiliary continuous order parameter $\phi ({\bf
r})$ that couples to the evolution of the thermal or solutal
field. The phase field interpolates between the solid and liquid
phases, attaining two different constant values in either phase,
(e.g., $\pm 1$) with a rapid transition region in the vicinity of the
solidification front.  The level set of $\phi ({\bf r})= 0$ is
identified with the solidification front, and the subsequent dynamics of
$\phi$ are designed to follow the evolving solidification  front
in a manner that reproduces the Stefan problem 
\cite{LanS86,Cag86a,Col85,War95,Whe92,Kar94,Eld94,Whe96,Kob93,Pro96,Wan96}.

The price to be paid for the convenience of the order parameter
is the introduction of a new length scale $W$ which represents 
a boundary layer over which the order parameter changes sign. This distance
is referred to as the interface width, and does not appear in the Stefan 
problem. As such, one requirement of the phase-field model is to recover the
Stefan limit in a manner that is independent of the interface 
width as $W$ approaches some appropriate limit. 
considerable work has been done to relate $W$ to various parameters of 
the phase-field model  in order to establish a mapping between the phase-field model 
and the Stefan problem \cite{Cag86a,Kar95,Fab97}. While the formal nature of 
these mappings does not seem to be very sensitive to the precise form of the phase-field
model\cite{Fab97}, different asymptotic limits of the
phase-field parameters can lead to widely varying complexity in
the numerical implementation.

The introduction of the interface width $W$ makes the phase-field
model prohibitively expensive to simulate numerically for large 
systems, since the grid spacing must be small enough everywhere that the
phase-field model converges to the the sharp interface
limit\cite{Cag86a,Kar95}.  Caginalp and Chen \cite{Cag92} showed
rigorously that the phase field model converges to the sharp
interface limit when the interface width (and hence the grid
spacing) is much smaller than the capillary length.
While this result is necessary to establish that the phase-field
model does map onto the Stefan problem, the parameter values
required to realize the asymptotic limit can be computationally intractable.
Experimentally, the physical sizes required to contain realistic  
microstructures can be many times the size of the thermal diffusion 
length, which in turn can be orders of magnitude greater than $W$
Thus, since $\Delta x_{\rm min} < W$, computing in the limit of
a $W \rightarrow 0$ does not allow one to simulate experimental systems.

Recently Karma and Rappel\cite{Kar95} presented a different
asymptotic analysis performed in powers of the ratio of the interface
width to the diffusion length $\alpha/V_n$, taken to be equal in both phases.  
Their procedure offers two computational 
advantages. The first is that is allows one to simulate the 
phase-field model with zero interface kinetics, {\it without}  
the need to make $W \rightarrow 0$. Specifically, this limit, 
as well as a non-zero kinetics limit, can be simulated 
with an interface width $W$ (and hence the grid spacing) of 
order the capillary length,  a much more tractable regime.  
Simulating solidification microstructures in the limit of 
zero interface kinetics is important because most experiments 
performed at  low undercooling in materials such as succinonitrile are 
in this limit \cite{Gli84}.  Karma and Rappel tested 
their asymptotics by comparing their simulations to the 
results of microscopic solvability theory, finding excellent 
agreement down to dimensionless undercoolings as low as 0.25.  

A recent extension of Karma and Rappel's analysis by Almgren \cite{Alm97} 
also promises to allow
similar asymptotics to be performed on a two-sided model of
solidification \cite{Alm97}, i.e. when the diffusivities in the 
solid and liquid differ, relevant in the study of
directional solidification of binary mixtures.

The theory of level sets \cite{Che97,Mer98} has also recently 
re-emerged as another effective tool that shows great potential 
in modeling dendritic growth. While related to the phase-field model, 
level-set theory does not require the presence of a thin interfacial 
with $W$, thus greatly reducing the stringent grid requirements posed by 
conventional phase-field models. To date, however, level set 
methods have not been benchmarked with solvability theory or other 
theoretical prediction for Stefan problems.

While expanding the horizon of solidification modeling, 
phase-field modeling has still been limited to
small systems sizes, even when solved by 
adaptive algorithms \cite{Bra97}.  The main problem 
is the presence of an interface region
with a minimal length scale that must be resolved.
For a typical microstructures grown at dimensionless undercooling
of $0.1$ or less, the ratio of the system size to this
minimal grid spacing can be greater that $2^{17}$.  With this
restriction most numerical methods will naturally fail. 
What is needed to go beyond this limitation is an effective adaptive 
technique\cite{Bra97,Nee96,Sch96,Pro98} which dynamically coarsens the grid 
spacing away from the front.

In this paper we present a new, computationally efficient
adaptive-grid algorithm for solving a class of phase-field
models suitable for the study of phase-boundary evolution.  We 
study dendritic solidification modeled using two coupled 
fields, one for the order parameter and
the other for the thermal field.  Our algorithm effectively combines
and implements ideas of adaptive-mesh refinement with ideas of 
dynamic data structures, allowing us to enlarge the window of large-scale 
solidification modeling.  

The outline of this paper is as follows: In section one we
introduce the physical model to be examined, summarizing its
properties and its various limits. In section two we present a
detailed description of our algorithm.  In
section three we present results on CPU-scalability of our
algorithm and examine issues of grid convergence and grid
anisotropy on our solutions. In section four we present results
of dendritic growth with and without the presence of anisotropy
in the surface energy. We show that for high undercooling, 
dendrites grown with our method converge to tip speeds in agreement 
with microscopic solvability theory. At low undercooling, however, we
do not find agreement with steady state solvability theory, owing to long-lived 
transients in the thermal field evolution. In section four we 
conclude and discuss our results.

\section{The Phase-Field Model}

We model solidification using a standard form of 
phase-field equations which  couple a thermal field to an order
parameter field $\phi$  via a double well potential
\cite{Kar95,Cag86a}.  We begin by rescaling the temperature
field $T$ by $U=c_P(T-T_M)/L$, where $c_P$ is the specific heat
at constant pressure, $L$ is the latent heat of fusion and $T_M$
is the melting temperature. The order parameter is defined by
$\phi$, where we define $\phi=1$ in the solid phase and
$\phi=-1$ in the liquid phase. The interface is defined by
$\phi=0$.  We rescale time throughout by $\tau_o$, a time
characterizing atomic movement in the interface. Length is rescaled by
$W_o$, a length characterizing the liquid--solid interface. With
these definitions, the model is written as 
\begin{eqnarray}
&&\frac{\partial U}{dt} = D \nabla^2 U + \frac{1}{2}
\frac{\partial \phi}{\partial t} \label{phase-field}\\ 
\nonumber
&A^2(\vec{n})&  \frac{\partial \phi}{dt} = \nabla \cdot
(A^2(\vec{n}) \nabla \phi )  + g^{'}(\phi) - \lambda U
P^{'}(\phi) \\ \nonumber & + & \frac{\partial }{\partial x}
\left( |\nabla \phi|^2 A(\vec{n}) \frac{\partial
A(\vec{n})}{\partial \phi_{,x}} \right) + \frac{\partial
}{\partial y} \left( |\nabla \phi|^2 A(\vec{n}) \frac{\partial
A(\vec{n})}{\partial \phi_{,y}} \right), 
\end{eqnarray} 
where $D=\alpha \tau_o/W_o^2$ and $\alpha$ is the thermal
diffusivity.  The function $f(\phi,U;\lambda)=g^{\prime }(\phi) -
\lambda U P^{\prime}(\phi)$ is the derivative of the double-well
potential with respect to $\phi$ and couples the $U$ and $\phi$
fields via the constant $\lambda$. The primes on the functions 
$g(\phi)$ and $P(\phi)$ denote derivatives with respect to $\phi$.
Following Karma and Rappel \cite{Kar95}, anisotropy has been introduced in
Eqs.~(\ref{phase-field}) by defining the width of the interface
to be $W(\vec{n})=W_o A(\vec{n})$ and the characteristic time by
$\tau(\vec{n})=\tau_o A^2(\vec{n})$, with $A(\vec{n}) \in [0,1]$ given by
\begin{equation} 
A(\vec{n}) = (1- 3 \epsilon) \left[  1 +
\frac{4 \epsilon }{ 1 - 3 \epsilon} \frac{(\phi_{,x})^4 +
(\phi_{,y})^4 }{| \nabla \phi|^4}\right].  
\label{width}
\end{equation} 
The vector 
\begin{equation}
\vec{n}=\frac{\phi_{,x} \hat{x}+\phi_{,y} \hat{y} }{(\phi_{,x}^2 + \phi_{,y}^2)^{1/2} }
\label{normal} 
\end{equation} 
defines the normal to the contours of the $\phi$ field, where 
$\phi_{,x}$ and $\phi_{,y}$ are defined as the partial derivatives 
of $\phi$ with respect to $x$ and $y$.  The variable $\epsilon$ 
parameterizes the deviation of $W(\vec{n})$ from $W_o$ and represents the
anisotropy in the interface energy of the system. We note that this definition of
anisotropy is not unique \cite{Fab97}, but we expect results to
be similar for other definitions of anisotropy.

In simulating the phase-field model we adopt the point of view
that the order parameter field $\phi$ is a computational tool
whose main purpose is to eliminate front tracking. As such we
would like to simulate the model given by
Eqs.~(\ref{phase-field}) with $W_o$ as large as possible. At
the same time we would like the behavior of the
model outside the boundary layer defined by $\phi$ to describe
the Stefan problem as closely as possible. To this end, we
relate the parameters of the phase-field model according to 
Ref. \cite{Kar95}, valid in the asymptotic limit $W_o \ll \alpha/V_c$,
where $\alpha/V_c$ is the diffusion length and $V_c$ is a characteristic 
velocity of the front defined by $\phi$.  

The specific asymptotic limit we model is one where
the $U$-field satisfies
\begin{equation}
\frac{\partial U}{\partial t} = D \nabla^2 U
\label{stefan1} 
\end{equation}
everywhere away from the interface, while at the interface, the gradient of $U$ satisfies
\begin{equation}
V_{n} = D\left( \left. \frac{\partial U}{\partial \vec{n}} \right|_{\vec{x}_{\rm int}^{-}} -
\left. \frac{\partial U}{\partial \vec{n}} \right|_{\vec{x}_{\rm int}^{+}} \right),
\label{continuity} 
\end{equation}
where $V_{\rm int}$ is the velocity normal to the interface, denoted by 
$\vec{x}_{\rm int}$. The 
notation $\pm$ denotes the solid/liquid  side of the 
interface, respectively.  The description of the Stefan problem is completed 
by the Gibbs-Thomson condition  and the interface kinetics condition
\begin{equation} U(\vec{x}_{\rm int}) = -d(\vec{n}) \kappa - \beta(\vec{n}) V_n,
\label{gibb-thomson}
\end{equation} 
where $d(\vec{n})$ is the capillary length, $\kappa$ is the
local curvature and $\beta(\vec{n})$ is the interface attachment
kinetic coefficient,
all assumed to be in dimensionless form according to the above
definitions.  The 
capillary length is related to the parameters of Eqs.~(\ref{phase-field}) by
\begin{equation}
d(\vec{n}) =a_1 \frac{W_o}{\lambda} \left[ A(\vec{n}) + \partial_{\theta}^2
A(\vec{n}) \right] 
\label{cap_length} 
\end{equation} 
where $a_1=0.8839$ for the particular form of the free energy 
defined in Eqs.~(\ref{phase-field}) \cite{Kar95} and $\theta$ is
the angle between $\vec{n}$ and the $x$-axis.  The kinetic coefficient 
is given by 
\begin{equation}
\beta=\frac{a_1 \tau_o}{ \lambda W_o } \left[ 1 - \frac{\lambda a_2}{ D}\right]
\label{beta}
\end{equation}
where $a_2=0.6267$ for our choice of the free energy functional cite{Kar95}. 
One remarkable feature of Eqs.~(\ref{cap_length}) and (\ref{beta}) is that $W_o$, $\tau_o$ and $\lambda$ 
an be chosen to simulate arbitrary values
of $\beta$, for $W_o$ of order $d_o$. In particular, 
setting $\lambda=D/a_2$ allows us to
compute the phase-field model in the limit of the
Stefan problem \cite{Kar95}, where $\beta=0$. This is also 
an appropriate value for SCN, especially at low undercooling.

Equations~(\ref{cap_length}) and (\ref{beta}) for $\beta$ and
$d_o$ can be related to a wide class of free energies via the
parameters $a_1$ and $a_2$ \cite{Kar95}, which are 
related to integrals that depend on $g(\phi_o)$, $P(\phi_o)$ 
and $d \phi_o/dx$, where $\phi_o$ is the lowest order description 
of the order parameter field $\phi$ and and is a solution of the equation 
\begin{equation}
\frac{\partial^2 \phi_o}{\partial x^2} - \frac{ d g(\phi_o)}{d
\phi_o} =0.  
\label{phi_o} 
\end{equation} 

We also note that these asymptotics are a special case of a more general asymptotic 
analysis performed by Almgren \cite{Alm97}, which relates
the parameters of the phase-field model to those of the Stefan 
problem in the case of unequal diffusivities in the solid and liquid
phases. In this case, the asymptotics provides an additional set
of constraints on the admissible functions $P^{\prime }(\phi)$,
$g^{\prime}(\phi)$, and hence $a_1$ and $a_2$.

\section{The Adaptive-Grid Algorithm}

The main computational challenge of simulating
Eqs.~(\ref{phase-field}) involves resolving two competing length
scales: the lattice spacing $dx$ on which the simulation in performed
and the physical size of the system $L_B$.
Even with improved asymptotics, $dx$ must remain relatively
small, while $L_B$ must be extremely large in order to make
possible computations of extended solidification
microstructures. Moreover, the main physics of solidification
(and the evolution of most phase-boundary problems) occurs around 
an interface whose area is much smaller than the 
full computational domain.  Near this interface the order parameter 
varies significantly, while away from the interface variations 
in $\phi$ are small. Meanwhile, the thermal field $U$ extends
well beyond the interface and has much more gradual variation in 
its gradients, permitting a much coarser grid 
to be used to resolve $U$.  The most obvious manner to 
overcome this problem is to use a method that places a high density 
of grid points where the interface of $\phi$ or $U$ varies most
rapidly and a much lower grid density in other regions.
Furthermore, the method must dynamically adapt the grid to follow the
evolving interface \cite{Bra97,Nee96,Sch96,Pro98}, while at the same time 
maintaining a certain level of solution quality.

We solve Eqs.~(\ref{phase-field}) using the Galerkin finite
element method on dynamically adapting grids of linear,
isoparametric quadrilateral and triangular elements. The grid is
adapted dynamically based on an error estimator that utilizes
information from both the $\phi$ and $U$ fields.  We wrote our code in FORTRAN 90 (F90),
taking advantage of the efficiency of FORTRAN 77 while using 
advanced C-like features, such as data structures, derived data 
types, pointers, dynamic memory allocation and modular design
to conveniently adapt the grid and the solution fields.

In the broadest sense, our algorithm performs functions that can
be divided into two classes. The first deals with the
establishment, maintenance  and updating of the finite element
grids, and the second with evolving $\phi$ and $U$ on these grid,
according to Eqs.~(\ref{phase-field}). We presently describe 
these classes, the adaptive grids, the finite element 
procedure, and the interconnections of these processes.

\subsection{The Finite Element Grids}

\begin{figure}[h] 
\begin{center}
\leavevmode\mbox{\epsfxsize=2.4in\epsfbox{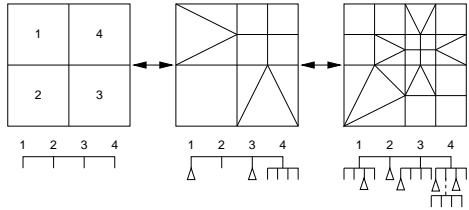}}
\end{center} 
\caption{An illustration of the quadtree element data structure.  
The first frame shows an element, and four child elements. 
Splitting of one of the children and one its children is shown, 
along with the branch evolution of the quadtree. Branches with triangles
indicate square elements which are bridged with triangular or rectangular 
elements.
}
\label{quadtree2}
\end{figure}

The first class of functions in our algorithm centers around
maintaining a grid of finite elements on a data structure
known as a {\it quadtree} \cite{Dev87,She88,Pal96} 
which replaces the standard concept of a uniform grid as a 
way of representing the simulational  grid.  The quadtree is a 
tree-like structure with branches up to a prespecified level. 
Branches of the quadtree are themselves data structures that contain
information analogous to their parent, from which they branched, 
but one level down.  Fig.~\ref{quadtree2} illustrates the structure of a quadtree as
well as the relation between elements at different levels of refinement. 
Every  entry on the quadtree contains information pertaining to 
a $4$-noded isoparametric quadrilateral finite element.  This information
includes the following: \\

\noindent $\bullet$ values of $\phi$ and $U$ at the four nodes \\

\noindent $\bullet$ the nodal coordinates of the element \\

\noindent $\bullet$ the level of refinement of the element on the quadtree \\

\noindent $\bullet$ the value of the current error estimate \\

\noindent $\bullet$  The element number, which contains information 
about the coordinates of the element and its level of refinement\\

\noindent $\bullet$ an array mapping the element's four 
    nodes onto the entries of a global solution array \\

\noindent $\bullet$ pointers to the element's nearest neighbors sharing a common 
edge and at the same level of grid refinement \\

\noindent $\bullet$ a variable that determines whether or not an element contains 
    further sub-elements which we term {\it child} elements \\

\noindent $\bullet$ pointers to an element's child elements \\ 

\noindent $\bullet$ a pointer to the {\it parent} element from 
which an element originates \\

The elements of the quadtree can be refined by splitting 
into four child elements, each sharing the same 
parent element one level down on the quadtree and each with 
its own unique information, as outlined above. A parent element and 
it's four child elements are referred to as a {\it family}.  
Refinement produces a finer mesh within the confines
of the original parent grid by bisecting each side, as shown in 
Fig.~(\ref{quadtree2}).
Unrefinement, which  consists of fusing the four child elements back into the parent, 
has the opposite effect, locally creating a coarser mesh. Both refinement and unrefinement proceed via 
dynamic memory allocation, making our code scalable. We note that unrefinement
can occur only if the child elements do not possess further child of their own.
Also, in order to avoid having regions of very different refinement bordering each other, 
we impose the restriction that any two neighboring quadrilateral elements may 
be separated by no more than one level of refinement (see Fig.~(\ref{quadtree2})).
We define the level of refinement of an element by $l_e$ such that a uniform grid 
at a refinement level $l_e$ would contain $2^{l_e} \times 2^{l_e}$
grid points in a physical domain $L_B \times L_B$.

Special cases where an element has no children, a missing neighbor, or no 
parent are handled by null pointers. The latter case occurs only for the root
of the quadtree.

All elements at a given level of refinement on the quadtree are
``strung" together by a linked-list of pointers, referred to as
the {\it $G$-lists}. There are as many $G$-lists as there are levels of 
refinement in the quadtree. Each pointer in the $G$-list points
to (accesses) the location in memory assigned to one element 
of the quadtree. The purpose of the $G$-list is to allow traversal 
of the quadrilateral elements sequentially by level, rather than 
by recursively traversing quadtree from the root down, 
a procedure which is memory intensive and relatively slow.

Alongside the main grid of quadtree elements, the code maintains two
independent grids representing special linear isoparametric triangular and
rectangular elements. These elements are used to connect the extra 
nodes that arise when two or more quadrilateral elements of differing 
refinement levels border each other.
These element types are referred to as
{\it bridging elements}. They are maintained as two linked-lists of
derived data types, one containing information about  triangular elements 
and the other rectangular. Elements of both these grids  include 
the following information:  \\

\noindent $\bullet$ the values of $\phi$ and $T$ 
    at the three nodes (four for rectangles) of the element \\

\noindent $\bullet$ the nodal coordinates \\

\noindent $\bullet$ node numbers that map the element's nodes 
onto the global solution array \\

\begin{figure}[h] 
\begin{center}
\leavevmode\mbox{\epsfxsize=2.4in\epsfbox{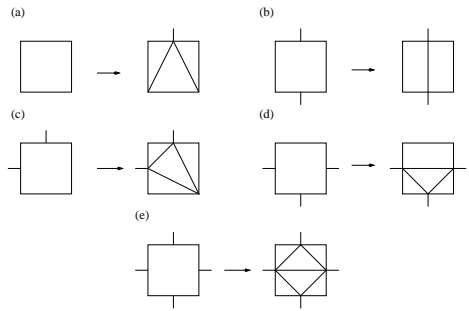}}
\end{center} 
\caption{Illustration of all possible 
configurations requiring completion with triangular and/or
rectangular elements.
}
\label{tri_recs_configs}
\end{figure}

\noindent The types of bridging triangles and rectangles that can occur are enumerable
and shown in Fig.~(\ref{tri_recs_configs}). 

The main set of operations performed on the grids described above 
concern refinement of the finite element mesh as a whole. 
The refinement process is performed only on the quadrilateral mesh.
The triangular and rectangular grids are established after this process 
is completed (see Fig.~(\ref{quadtree2})).  To refine the grid the code traverses the 
elements of the quadtree, refining (unrefining) any element 
whose error function, discussed below, is above (below) a critical
value $\sigma_h$($\sigma_l$). We also note that fusion of four quadrilateral elements
can occur only if all four of its children's error functions are below the
critical value $\sigma_l$, where $\sigma_l < \sigma_h$.  
We found that if $\sigma_l = \sigma_h$ the grid sets into 
oscillations, where large numbers of elements become alternatively refined at one
time step, then unrefined at the next.

\begin{figure}[h] 
\begin{center}
\leavevmode\mbox{\epsfxsize=2.4in\epsfbox{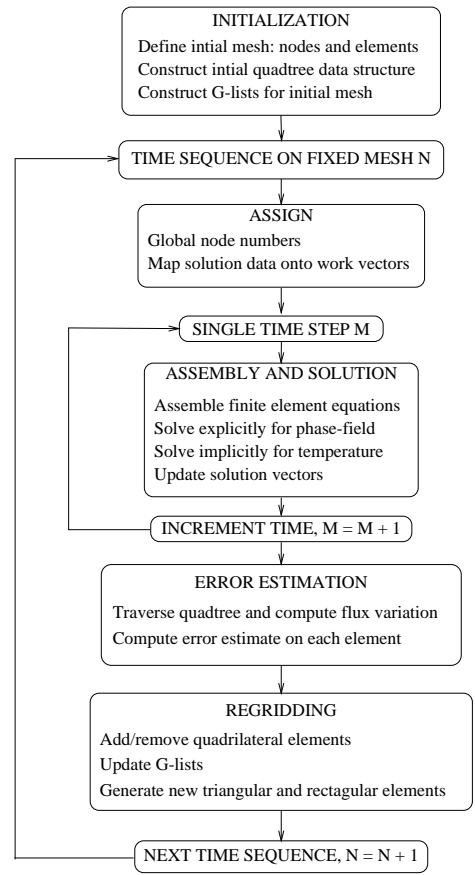}}
\end{center} 
\caption{A flow chart illustrating the algorithm 
program modules.
}
\label{flow_chart}
\end{figure}

The processes described thus far are grouped into modules that
encapsulate various related tasks, and which can cross-reference
each other's data and instructions. The module highest up in the hierarchy 
contains the definition of the quadtree data
structure and routines that construct the initial uniform grid,
refine and unrefine individual quadrilateral elements, and set
the initial conditions.  Another module constructs the G-lists.
It contains  routines that construct the initial G-list from
initial uniform quadtree data structure, as well as add or delete
element pointers from the $G$-list as elements are created or deleted 
from the quadtree. Another module accessing both the previous ones'
data structures has the role of creating the triangular and rectangular 
element grids. It contains definitions for creating triangular and 
rectangular elements data structures and routines that search the quadtree, building
the linked lists of triangles and rectangles that make up these grids. The main
program is contained in its own module and contains the driver
program that creates the initial grids, $G$-lists and 
triangular and rectangular element types. The driver program 
also sets into motion the final link in the simulation, which 
evolves $\phi$ and $U$ and periodically adapts the dynamic grid by calling procedures 
described above. A flowchart of these processes is shown in Fig.~(\ref{flow_chart}).

\subsection{The Finite Element Formulation}

The integration of Eqs.~(\ref{phase-field}) is done by the final
module in the code.  This module performs  four main processes: \\

\begin{enumerate}
\item Maps the internal element node numbers to the indices of 
a global solution vector. The $\phi$-field is mapped onto the 
odd numbers, ($1,3,5, \dots$), while $U$ is stored on the even 
numbers of the global solution vector ($2,4,6, \dots$) \\

\item Advances the $U$ and $\phi$ field-vectors by $N_r$ time steps on 
the finite element grids defined above \\

\item Calculates an error function for each element of the quadtree, 
based on error estimate of the quadrilateral elements \\

\item Invokes routines in the modules described above to refine the 
grid according to this error estimator \\

\end{enumerate}

Steps (1)-(4)  are repeated  until a sufficient time evolution of 
the microstructure is established.  The variable $N_r$ is 
set such that the interface remains within the regions 
of fine mesh between regriddings, which we typically choose to be 
100 time steps.  Step (1) 
involves searching all elements, and their neighbors, and assigning each node 
a unique number, that will have a counterpart on a global solution 
vector. 

The finite element discretization of Eqs.~(\ref{phase-field}) is done using Galerkin's 
weighted residual method \cite{Coo89}. The method begins by assuming that
$\phi$ and $U$ are interpolated within an element as 
\begin{equation}
\phi^e = \sum_{i=1}^{N} \phi_i^e N_i(x,y)  \hspace{0.5cm}
U^e = \sum_{i=1}^{ N} U_i^e N_i(x,y) 
\label{interpol_fields}
\end{equation}
where $\phi_i^e$ and $U_i^e$ are the field values at the 
$N$ nodes of the element $e$, and their interpolated values in its interior.
The functions $N_i(x,y)$ are standard 
linear interpolation functions appropriate to the element being used 
\cite{Zie87}, and satisfy 
\begin{equation}
N_i(x_j,y_j)=\delta_{i,j}, 
\label{shape_funcs_nodes}
\end{equation}
where $\delta_{i,j}$ is the Kroneker delta.
Rewriting the differential equations for $\phi$ 
in Eqs.~(\ref{phase-field}) as $L_{\phi} \phi =0$, and 
of the $U$-equation as $L_U U=0$, the Galerkin method requires that 
\begin{eqnarray}
&&\int_{\Omega_e} N_i(x,y) L_{\phi} \phi^e(x,y)  dx dy = 0 
\label{Gal_state} \\
\nonumber
&&\int_{\Omega_e} N_i(x,y) L_{U} U^e(x,y)  dx dy = 0,
\end{eqnarray}
for $i=1,2,3,\dots, N$, where $\Omega_e$ represents the area of an element 
$e$.  Substituting Eqs.~(\ref{interpol_fields}) 
into Eqs.~(\ref{Gal_state}), we obtain two linear algebraic 
equations for $\phi_i$ and $U_i$, $i=1,2,3, \dots, N$
in the element $e$.  

We next define $\{\Phi\}^e=(\phi_1,\phi_2,\phi_3, \cdots, \phi_{N})^T$
and $\{U\}^e=(U_1,U_2,U_3, \cdots, U_{N})^T$, where the superscript $T$ denotes 
transpose, making $\{\Phi\}^e$ and $\{U\}^e$ column vectors.  
The linear algebraic statement of the finite element form of 
Eqs.~(\ref{phase-field}) then  becomes
\begin{eqnarray}
&&[\hat{C}](\{ \phi \}^e) \frac{d \{ \phi^e\}}{dt}  = \left([M]+[E]\right) \{\phi\}_{n}^e + \{F; \lambda\}^e
\label{Mat_eqs}\\
\nonumber
&&[C] \frac{d \{U\}^e}{dt}   =D [A] \{U\}^e + \frac{1}{2} [C] \frac{d \{ \phi\}^e }{dt},
\end{eqnarray}
where the matrices $[C]$, $[\tilde{C}]$, $[A]$, $[M]$ and $[E]$ and 
the vector $\{F; \lambda\}^e$ are given by 
\begin{equation}
[C] = \int_{\Omega_e} [N]^T [N] dx dy, 
\label{C_mat}
\end{equation}
\begin{equation}
[\hat{C}] = \int_{\Omega_e} [N]^T [N] A^2(\theta (\phi^e))  dx dy, 
\label{Chat_mat}
\end{equation}
\begin{equation}
[A] = - \int_{\Omega_e} \left( [N]^T [N_x] + [N]^T [N_y] \right) dx dy, 
\label{A_mat}
\end{equation}
\begin{equation}
[M] = - \int_{\Omega_e} \left( [N]^T [N_x] + [N]^T [N_y] \right) A^2(\theta(\phi^e ))  dx dy, 
\label{M_mat}
\end{equation}
\begin{equation}
[E] = - \int_{\Omega_e} \left( [N]^T [N_x] - [N]^T [N_y] \right) 
A(\theta(\phi^e)) \omega(\theta(\phi^e)) dx dy,
\label{E_mat}
\end{equation}
\begin{equation}
\{F;\lambda\}^e = \int_{\Omega_e} [N]^T f(\phi^e, U^e ;\lambda) dx dy,
\label{F_mat}
\end{equation}
where $[N_x]$, $[N_y]$ denote the partial derivatives of the vector of shape functions 
with respect to $x$ and $y$, respectively. The function $A$ is just Eq.~(\ref{width})
rewritten in terms of the angle $\theta$ that the normal to the interface makes
with the $x$-axis. Specifically, defining 
\begin{equation}
\tan \theta({\phi}^e) = \frac{\partial \phi_{,y}^e}{\partial \phi_{,x}^e}.
\label{theta_def}
\end{equation}
then 
\begin{equation}
A(\theta({\phi}^e))=(1-3\epsilon) \left[ 1 + \frac{4 \epsilon}{1 - 3\epsilon}
\frac{(1+ \tan^4 \theta)}{(1+ \tan^2 \theta)^2}   \right]
\label{alternate_A}
\end{equation}
while $\omega(\theta)$ is proportional to the derivative of $A(\theta)$, 
and is given by 
\begin{equation}
\omega(\theta({\phi}^e)) = 16 \epsilon 
\frac{\tan \theta (1 - \tan^2 \theta)}{(1+ \tan^2 \theta)^2}, 
\label{W_prime}
\end{equation}

We use a lumped formulation for the matrices $[C]$ 
and $[\hat{C}]$ \cite{Coo89}. In this procedure, the row vector 
of shape functions, $[N]$ in Eq.~(\ref{C_mat}) is replaced by the 
identity row vector $[I]=[1,1,1,\cdots]$. The resulting matrix $[C]$
then consists of identical columns, each of which contains the element 
$N_i(x,y)$ in the position of the $i^{\rm th}$ row.  A lumped 
term is then defined as a diagonal matrix whose entries take 
on the value 
\begin{equation}
L_c= \frac{1}{4} \sum_{i=1}^{\rm nodes} \int_{\Omega_e} N_i(x,y) dx dy.
\label{lump_def}
\end{equation}
The use of a lumped matrix for $[C]$ allows us to assemble a diagonal 
matrix for the left hand side Eqs.~(\ref{Mat_eqs}), stored as a one-dimensional 
vector rather than two-dimensional matrices that would be required if we 
used the consistent formulation for  the assembly of the $[C]$ matrices.
Indeed, microstructures evolving at low undercooling can 
produce interfaces with over $2 \times 10^5$ elements, making the storing of 
$2 \times 10^{10}$ matrices impossible.

The global $\{\phi\}$ (obtained after assembly of the element 
equations in field in Eqs.~(\ref{Mat_eqs})) is time-stepped using
using a forward difference (explicit) time scheme.
For each time step of the $\phi$ field,  the global $U$ field is then 
solved iteratively using a Crank-Nicholson scheme. 
Convergence of $\{U\}_{n+1}$ is obtained in a few iterations.

\subsection{The Error Estimator}

Regridding is based on an error estimator function, which is
obtained following Zienkiewicz and Zhu \cite{Zie87}, based on
the differences between calculated and smoothed
gradients of the $\phi$ and $U$ fields. Specifically, we define
the {\it composite field} 
\begin{equation} 
\Psi = \phi + \gamma U \label{composite_def} 
\end{equation} 
where $\gamma$ is a constant. 
We discuss the selection of $\gamma$ in more detail below. This 
definition allows us to regrid according the requirements of both 
the $\phi$ and $U$ field, as opposed to using only gradients of the 
$\phi$-field in establishing  the grid \cite{Bra97}.
Since it is $\phi$ and $U$ that are being calculated, and not
their gradients, we do not expect the gradient of $\Psi$ to be
continuous across element boundaries, due to the order
of the interpolation used. Thus we expect the difference between 
the calculated and smoothed (continuous across element boundaries) 
gradients to provide a reasonable estimate of error. This method 
appropriately meshes regions of both steep gradients and regions 
where the $\phi$ and $U$ fields change rapidly.

We define the error estimator function $\vec{e}$ as 
\begin{equation} \vec{e} = \vec{q}_{\rm s} -\vec{q}_{\rm c} \label{error_def} 
\end{equation} 
where $\vec{q}_{\rm c}$ and $\vec{q}_{\rm s}$ are the
calculated and smoothed gradients of $\Psi$ respectively. 
The smoothed gradients are  calculated to be continuous across element
boundaries. To determine $\vec{q}_{\rm s}$ we assume it
to be interpolated in the same way as the $\phi$ and $U$ fields, namely
\begin{equation} \vec{q}_{\rm s} = [N]\{Q^s\}
\label{q_smooth_int} 
\end{equation} 
where $[N]$ is the row vector of element shape functions, 
and $\{Q^s\}$ is a $4 \times 2$ matrix whose columns represent the 
nodal values of fluxes of $\Psi$ in the $x$ and $y$ direction, respectively. 
To find $\{Q^s\}$ we use Galerkin's  method, minimizing the weighted
residual 
\begin{equation} \int_{\Omega_e} [N]^T \vec{e}
d\Omega_e=\int_{\Omega_e}[N]^T([N]\{Q^s\}-\vec{q}_{\rm c})d\Omega=0
\label{error_residual} 
\end{equation} 
The calculation is simplified by lumping the left hand side 
of Eq.~(\ref{error_residual}),  leading to 
\begin{equation} \left( \int_{\Omega_e} [N]^T [{\bf 1}] d\Omega \right)\{Q^s\}
= \int_{\Omega_e} [N]^T \vec{q}_{\rm c} d\Omega, 
\label{error_matrix} 
\end{equation} 
where $[{\bf 1}]=[1,1,1,\cdots,N]$. 
Assembling Eq.~(\ref{error_matrix}) for all quadrilateral 
elements yields an equation for the smoothed gradients $\{Q\}^g$ of  
the global field $\Psi$, at all element nodes, of the form 

\begin{equation}
[D]\{Q\}^g = {b},
\label{error_matrix_short}
\end{equation}
where $[D]$ is a diagonal matrix, due to ``mass" lumping, and $\{Q\}^g$ 
is a $N \times 2$ matrix for the global, smoothed flux.

For the actual error updating on the elements of the quadtree we 
used the normalized error defined by 
\begin{equation}
E_e^2 =\frac{ \int_{\Omega_e} |( \vec{q}_{\rm s} - \vec{q}_{\rm c} )|^2 } 
{ \sum_e \int_{\Omega} | \vec{q}_{\rm s} |^2 }.  
\label{norm_error}
\end{equation} 
The domain of integration $\Omega$ in the denominator denotes the entire domain of the problem.
Thus $E_e^2$ gives the contribution of the local 
element error relative to the total error calculated over the entire grid.

\begin{figure}[h]
\begin{center}
\leavevmode\mbox{\epsfxsize=2.4in\epsfbox{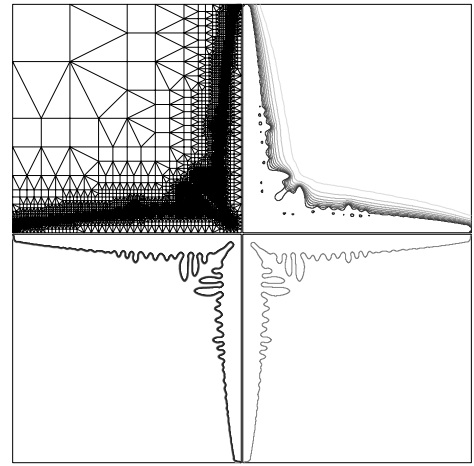}}
\end{center}
\caption{A dendrite grown using the adaptive-grid method for
$\Delta=0.7$, $D=2$, $\epsilon=0.05$. Clockwise,
beginning at the upper right the figures show contours of the $U$-field,
the contour $\phi=0$, contours of the $\phi$-field and
the current mesh.
}
\label{dendrite_picture}
\end{figure}

Fig.~(\ref{dendrite_picture}) shows a snapshot at $10^5$ time steps into the 
a simulation of a thermal dendrite computed with our algorithm. The figure shows
$\phi$ and $U$ as well as the  current grid. 
The dendrite is four-fold symmetric, 
grown in a quarter-infinite space, initiated by a small
quarter disk of radius $R_o$ centered at the origin.  The order
parameter is defined on an initially uniform grid to be its
equilibrium value
$\phi_o(\vec{x})=-\tanh((|\vec{x}|-R_o)/\sqrt{2} )$ along the
interface.  The initial temperature decays exponentially from
$U=0$ at the interface to $-\Delta$ as $\vec{x} \rightarrow
\infty$. 
The parameters set for this simulation are $\Delta=0.70$, $D=2$, $dt=0.016$
and $\lambda$ chosen to simulate $\beta=0$.  The system size is
$800 \times 800$, with $\Delta x_{\rm min}=0.4$, and about half
of the computational domain in each direction is shown.  Sidebranching is evident,
and arises due to numerical noise. This simulation was completed in 
approximately 15 cpu-hours on a Sun UltraSPARC 2200 workstation.

\section{Scalability and Convergence Properties of the Adaptive-Grid Algorithm}

In this section we present results that illustrate the convergence
properties of solutions of Eqs.~(\ref{phase-field}) computed with our algorithm, the
effect of grid-induced anisotropy of the adapting mesh, and the  speed
increase obtained by using an adapting grid.

\subsection{CPU-Performance}

\begin{figure}[h]
\begin{center}
\leavevmode\mbox{\epsfxsize=2.4in\epsfbox{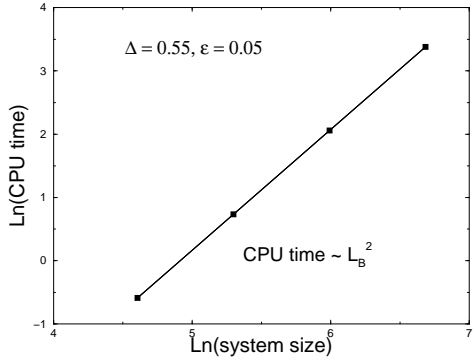}}
\end{center}
\caption{CPU time vs.~the system size, illustrating the
computing time for a dendrite to move through
the system of linear dimension $L_B$ using our adaptive mesh method.
}
\label{cpu}
\end{figure}

We examined the cpu-scalability of our algorithm as a function of 
system size by growing dendrites in systems of various linear dimension
$L_B$ and measuring the cpu time required for the dendrite
branches to traverse the entire system. Fig.~\ref{cpu} shows a
plot of these data for a dendrite grown at undercooling
$\Delta=0.55$ using the same parameters as in
Fig.~\ref{dendrite_picture}. The minimum grid spacing has 
been set to $\Delta x_{\rm min}=0.4$ in this data.
Fig.~\ref{cpu}, clearly shows that $R_t^a \sim L_B^2$.  This
relationship can be obtained analytically as follows. 

The number of calculations performed, per simulation time step, is
proportional to the number of elements in the grid. This relationship 
is set in turn by the arclength of the interface being simulated 
multiplied by the diffusion length $D/V_n$. This product 
defines the arclength over which the highest level of refinement occurs. 
For a needle-like dendrite, the arclength is approximately 
$L_B$. Moreover, 
since the dendrite tip moves at a constant velocity $V_n$, then 
\begin{equation} R_t^a = \left[
\frac{R_o^a D }{V_n^2 \Delta x_m^2} \right] L_B^2,
\label{cputime} 
\end{equation} 
where $R_o^a$ is a constant
that depends on the details of the implementation of the algorithm used to 
evolve Eqs.~(\ref{phase-field}).  The cpu time needed to compute the traversal 
time on a uniform grid, $R_t^u$, is found, by the same analysis, to be
\begin{equation} R_t^u
=\left[ \frac{R_o^u }{ V_n \Delta x_m^2} \right] L_B^3.
\label{time_uniform} 
\end{equation} 
where $R_o^u$ also depends on the implementation but is likely  to be smaller
than $R_o^a$. 
Thus, comparing our method with simulation on a uniform grid we obtain 
\begin{equation}
\lim_{L_B \rightarrow \infty} R_t^a/R_t^u =  \frac{1}{L_B}.  
\end{equation}
For larger systems, the adaptive scheme should always provide faster
CPU performance regardless of implementation. Indeed, any method
that uses a uniform grid of any sort, will eventually be limited
by memory requirements as $L_B$ becomes 
large.  The arguments leading to Eq.~(\ref{cputime}) can also be 
generalized to any problem of evolving phase boundaries, always yielding 
the conclusion that cpu time scales with arclength
in the problem being considered.  We note that when interface
convolutions become  of order $\Lambda \sim \Delta x_{\rm min}$, fine-grid 
regions separated by less than $\Lambda$ will merge and the number of
elements will stop growing locally. This makes the simulation of
fractal-like patterns feasible as the arclength of the interface is
bounded from above by $L_B \times L_B$. Finally, we note that adaptive gridding
would especially improve the cpu performance of problems similar to spinodal
decomposition, where the total interface decreases with time.

\subsection{Induced Lattice Anisotropy}

We tested the effective anisotropy of our dynamically adapting
lattice in two independent ways. The first follows the method
outlined by Karma\cite{Kar95}. We fix the 
temperature far from the interface to be constant $T_{\infty }$ everywhere, 
initially setting it to a critical
value at which the isotropic surface energy just balances  the
bulk free energy. For a specified background temperature, the crystal 
will only grow if its radius is greater than a critical value $R_o$.
The radius $R_o$ can be related to the background temperature
through the total Gibbs free energy of the system, given by 
\begin{equation} \Delta G =- \pi r^2\frac{L \Delta T}{T_M} + 2\pi
r \sigma, 
\label{Gibbs_free_energy} 
\end{equation} 
where $L$ is the latent heat of fusion, $\Delta T=T_m-T_{\infty}$, where $T_m$ is
the melting temperature, $T_{\infty}$ is the temperature far away from the interface,
and $\sigma$ is the surface
tension.  Minimizing $\Delta G$ with respect to $r$ yields $R_o$ as a function 
of $\delta T$ as 
\begin{equation} 
R^* = d_o/\Delta T \label{critical_radius}, 
\end{equation} 
where $d_o$ is the capillary length defined as $d_o=2 \sigma T_M/L$.

One finds an equilibrium shape of the interface when the 
background temperature field  $\Delta T$ (written in terms 
of $U$) is adjusted dynamically so as to maintain the velocity 
of the interface at zero as measured long the $x$-axis. 
Thus, $\Delta T$ is increased if the velocity decreases, and decreased if it grows.
The effective anisotropy is inferred by fitting the computed
interface to an equation of the form 
\begin{equation}
R(\theta) = R_o(1 + \epsilon_{\rm eff} \cos \theta),
\label{anisotropy_func} \end{equation} 
where $R(\theta)$ is the radial distance from the center of the crystal to 
its interface and $\theta$ the
polar angle. The effective anisotropy $\epsilon_{\rm eff}$
represents the modification of the anisotropy $\epsilon$ due to
the grid.  Fig.~(\ref{aniso_int}) illustrates a
crystal grown to equilibrium using an input anisotropy
$\epsilon=0.04$. Using Eq.~(\ref{anisotropy_func}) we found
$\epsilon_{\rm eff}=0.041$, within 5\% of $\epsilon$. Similar
accuracy was found for $\epsilon=0.02, 0.03$ and  $0.05$.

\begin{figure}[h]
\begin{center}
\leavevmode\mbox{\epsfxsize=2.4in\epsfbox{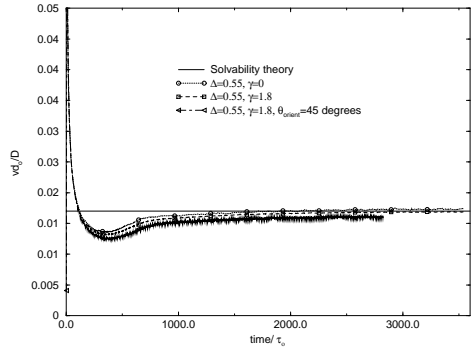}}
\end{center}
\caption{The time evolution of the tip-velocity of a dendrite
growing in the presence of surface tension anisotropy for
$\Delta=0.55$. Data is shown for the cases where the dendrite 
is moving in the $x$-direction with two grid layering patterns, 
and along the 45 degree line.  The horizontal solid line
represents the analytic prediction of microscopic solvability. 
}
\label{conv_vel0.55}
\end{figure}

We also tested for grid anisotropy by rotating the grid by $45$
degrees, which should represent the lowest accuracy for square
elements.  We compared the tip speed of dendrites grown in this
direction to that of dendrites whose principal growth direction
is along the x-axis.  Fig. ~\ref{conv_vel0.55} shows
the tip velocity for the case of a dendrite grown at
$\Delta=0.55$ ($\epsilon=0.05$, $D=2$, $\beta=0$, 
$dt=0.016$, $\Delta x_{\rm min}=0.4$) compared with the same case 
when growth occurs along the $x$-axis. The tip velocity approaches 
an asymptotic value that is within approximately  $5\%$ of the tip 
velocity computed when the anisotropy is aligned with the $x$-direction.

\subsection{Convergence and  Grid Resolution}

We tested the convergence of solutions as a function of the
minimum grid spacing $\Delta x_{\rm min}$. We used an
undercooling of $\Delta=0.55$, with $D=2$, $dt=0.016$, 
$\Delta x_{\rm min}=0.4$, 
and set $\lambda$ to simulate $\beta=0$. The parameter 
$\gamma=1.8$, which assured that 
regions of rapid change of $\phi$ and $U$ were always encompassed 
in the regions of highest grid resolution. We examined the tip
speed of a dendrite for $0.3 \le \Delta x_{\rm min} \le 1.6$,
finding relatively good convergence of the tip speed to
theoretical prediction of microscopic solvability theory
discussed above. Fig.~(\ref{rich}) shows the asymptotic
steady state tip velocity for each case, superimposed on the solid
line, which is the result of solvability theory for
$\Delta=0.55$. It is surprising that the solution
convergence is rather good even for $\Delta x_{\rm min} =1.6$.
We have found similar convergence properties for the case of
$\Delta=0.25$.  Specifically, using $\Delta x_{\rm min}=0.4$ and $\Delta
x_{\rm min}=0.78$ gives essentially identical results.

\begin{figure}[h]
\begin{center}
\leavevmode\mbox{\epsfxsize=2.4in\epsfbox{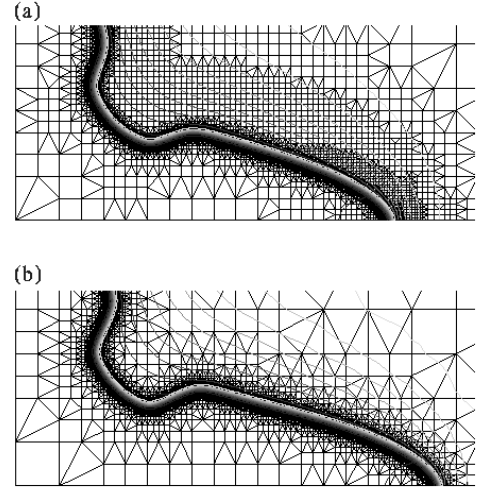}}
\end{center}
\caption{The finite element mesh around a dendrite branch 
growing at $\Delta=0.65$, showing the grid configuration 
for  (a) $\gamma=4$ and (b) $\gamma=0$. The grey-shaded lines 
represent isotherms ranging from $-0.65 \le U \le 0.02$.
}
\label{grid_layer}
\end{figure}

\begin{figure}[h]
\begin{center}
\leavevmode\mbox{\epsfxsize=2.4in\epsfbox{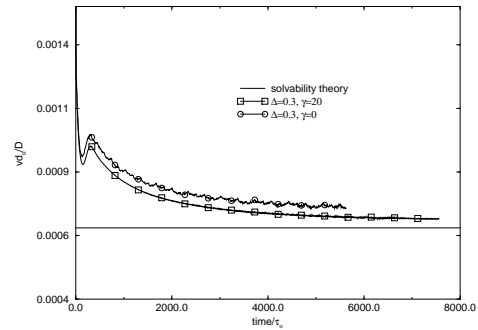}}
\end{center}
\caption{The tip-velocity of a dendrite
for $\Delta=0.3$. Data are shown for two grid layering patterns.
The horizontal solid lines represent the analytic prediction of
microscopic solvability.
}
\label{conv_vel0.3}
\end{figure}

The introduction $\gamma$ in the error function $\Psi$ gives us
the freedom to tune the degree to which the fine grid layering encompasses the thermal
field as well as the $\phi$ field.  Setting $\gamma=0$ leads to
a uniform-like mesh at the highest level of refinement 
{\it only} around the most rapidly changing
regions of $\phi$, while the $U$ field becomes encompassed in a
rather disorderly combination of quadrilateral and triangular
elements.  We found that this effect can increase the tip-speed 
error by several percent, as well as increase fluctuations in tip speed. 
Increasing $\gamma$ produces a smooth layering
of coarser uniform-like meshes  ahead of the $\phi$-field, corresponding
to region of large gradients in $U$.
Fig.~\ref{grid_layer} compares the mesh around the tip of a
dendrite grown at $\Delta=0.65$ for  $\gamma=0$ and
$\gamma=4$. The figure illustrates the gradual mesh layering
encompassing the thermal field for $\gamma=4$. In Fig.~(\ref{grid_layer}), 
$D=1$, $dt=0.016$, $\Delta x_{\rm min} =0.4$ and $\lambda$ is chosen to simulate
$\beta=0$. Fig.~\ref{conv_vel0.55} also shows the tip speed for 
$\Delta=0.55$ for the cases $\gamma=0$ and $1.8$, while Fig.~(\ref{conv_vel0.3}) 
shows the tip velocity for a dendrite grown at $\Delta=0.3$ with 
$\gamma=0$ and $20$, respectively. In Fig.~(\ref{conv_vel0.3}) $D=10$, 
$\Delta x_{\rm min}=0.4$, $dt=0.048$ and $\beta=0$.
In this case the higher value for $\gamma$ allows the tip velocity to approach 
within approximately 5\% of the solvability answer, as in Ref.~\cite{Kar95}. 
Raising $\gamma$ further does not produce any further changes in tip speed. 

\section{Dendritic Growth using Adaptive Gridding}

In this section we present results for two dimensional 
solidification with and without interface anisotropy.
We illustrate the robustness of our algorithm and use 
it, in particular, to investigate dendritic growth at 
low undercooling, presenting new results on dendrite tip-speed 
selection.

\subsection{Dendritic Growth without Surface Tension Anisotropy}

\begin{figure}[h]
\begin{center}
\leavevmode\mbox{\epsfxsize=2.4in\epsfbox{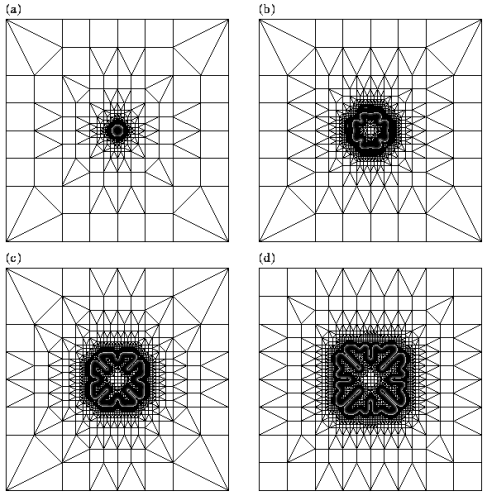}}
\end{center}
\caption{The evolution of a crystal growing without interfacial
anisotropy. The $\phi=0$ contours are shown, superimposed on
the finite element grids. Time advances from left to right, top 
to bottom.
}
\label{disk_growth_grids}
\end{figure}

When the anisotropy parameter $\epsilon$ in
Eqs.~(\ref{phase-field}) is set to zero, solidification 
proceeds without the emergence of any preferred direction.  In
this case it is well known that a seed crystal larger than 
a critical radius will eventually grow to 
become unstable to fluctuations, and  will
break into surface undulations via the Mullins-Sekerka 
instability \cite{Try96}. Fig.~\ref{disk_growth_grids} shows a series of
time steps in the evolution of a solidifying disk grown at
$\epsilon=0$, $\Delta=0.65$, $D=4$, and $\lambda$ set to
generate $\beta=0$, making $d_o=0.1385$. We use 11 levels of
refinement and an $800 \times 800$ system, making 
$\Delta x_{\rm min} =0.4$. For coarser meshes the
Mullins-Sekerka instability sets in sooner due to grid noise. As
$\Delta x_{\rm min}$ is made smaller, grid noise becomes smaller, and
one must wait longer for the true ``thermal noise" to set in and
make the crystal interface unstable.
The dynamically evolving grids are also shown.  
This figure clearly demonstrates  how the grid
creation scales with the arclength of the solidifying surface.

\subsection{Dendritic Growth with Surface Tension Anisotropy}

\begin{figure}[h]
\begin{center}
\leavevmode\mbox{\epsfxsize=2.4in\epsfbox{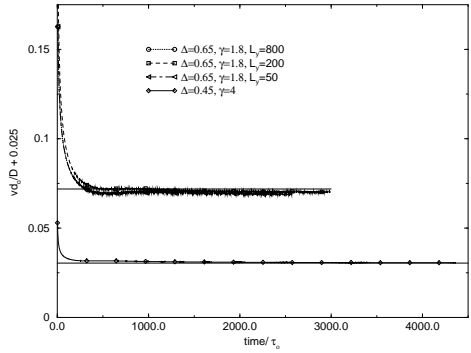}}
\end{center}
\caption{The time evolution of the dimensionless tip 
velocity for $\Delta=0.45$ and $0.65$. The horizontal 
lines represent solvability theory.  The $\Delta=0.65$
case includes data for three system sizes. The data 
have been shifted up by $0.025$ for clarity.
}
\label{sys_vel}
\end{figure}

\begin{figure}[h]
\begin{center}
\leavevmode\mbox{\epsfxsize=2.4in\epsfbox{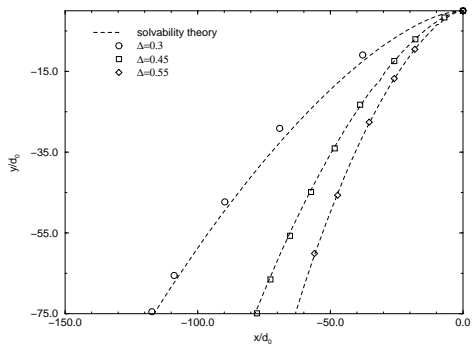}}
\end{center}
\caption{The asymptotic dendrite tip shapes for
$\Delta=0.3$, 0.45 and 0.55 (data points).
The  dashed lines are the shapes 
predicted by solvability theory.
}
\label{high_delta_shapes}
\end{figure}

When surface tension anisotropy is present a crystallizing disk
forms dendritic branches which travel along the symmetry axes of
the anisotropy,  driven by the anisotropy to a steady state tip
velocity \cite{Kes88,Ben83,Ben84,KessI84,Bre91,Pom91}.  As a verification of our
algorithm we measured tip-velocities and shapes for dendrites
grown at intermediate undercoolings. Fig.~\ref{sys_vel} shows
the dimensionless tip velocity ($V d_o/D$) versus time for
$\Delta=0.45$ and $0.65$ and $\epsilon=0.05$.
In Fig.~(\ref{sys_vel}) the dimensionless
diffusivities $D=3$ and $1$ and the dimensionless capillary length
are $d_o=0.186$ and $0.544$, respectively.  In both cases $\lambda$
has been set to simulate $\beta=0$ kinetics at the interface, while
$\gamma=4$ and $1.8$, respectively. These values of $\gamma$
are chosen so as to minimize grid-layering error. 
The solid horizontal lines represent the theoretical values
obtained from microscopic solvability theory.  In all cases the
converged velocities are within a few percent of the theoretical
prediction. 
The case of $\Delta=0.65$ includes data for three systems sizes. 
These results of system size are typical for intermediate $\Delta$,
showing a relatively rapid leveling to an asymptotic speed to
within a few percent of that predicted by solvability theory.
Fig.~(\ref{high_delta_shapes}) also shows a plot 
of the dendrite tip shapes produced by our simulations, superimposed
on the shapes predicted by solvability theory.

\subsection{Dendritic Growth at Low Undercooling}

\begin{figure}[h]
\begin{center}
\leavevmode\mbox{\epsfxsize=2.4in\epsfbox{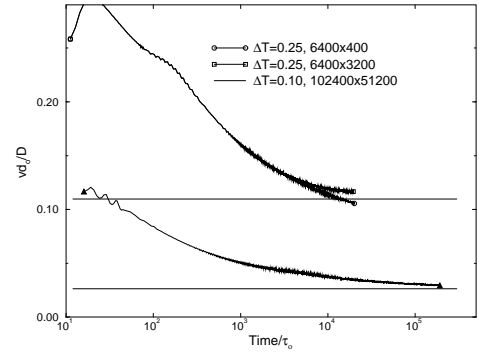}}
\end{center}
\caption{The time evolution of the tip-velocity for
undercooling $\Delta=0.25$ and $0.10$. The data 
have been shifted up by $0.025$ for clarity. 
}
\label{low_delta_vel}
\end{figure}

At lower undercooling we encounter significant finite-size effects
which cause the tip velocity to deviate from the solvability
prediction.  Fig.~\ref{low_delta_vel} shows the evolution
of the tip-velocity for $\Delta=0.25$ in two different system
sizes.  For a system of size $L_x=6400 \times L_y=400$, the
velocity goes to within a few percent of the solvability
prediction.  For a system size $6400 \times 3200$ the tip
velocity seems to settle close to a value that exceeds
the solvability prediction  by $8\%$.  This
effect is even larger at $\Delta=0.1$, also shown in
Fig.~(\ref{low_delta_vel}), where the tip speed approaches a
value about $3$ times larger than that predicted by
solvability theory.

To understand this finite-size dependence of tip velocity at low
undercooling, we note that at low $\Delta$, the thermal fields
of the two dendrite branches overlap, producing a thermal
envelope very different from that which emerges for the  single,
isolated dendrite branch  assumed in solvability theory.  At
large undercooling, each dendrite arm quickly outruns the
other's thermal boundary layer, and solvability theory should
apply, as is seen in Fig.~(\ref{dendrite_picture}) where
$\Delta=0.7$.  The conditions of solvability theory can be
better approximated at lower undercooling if simulations are
performed in a domain which is small in one direction. For the
simulation performed with $\Delta=0.25$ in a small box ($6400
\times 400$), the branch in the y-direction is extinguished by
its interaction with the wall and the velocity quickly
approaches the solvability prediction.  However, when both
branches are present, as in the simulation with $\Delta=0.25$ in
the larger box ($6400 \times 3200$), their interaction leads to
an increased tip-velocity because the dendrites are embedded in
a circular rather than parabolic diffusion field.  

\begin{figure}[h]
\begin{center}
\leavevmode\mbox{\epsfxsize=2.4in\epsfbox{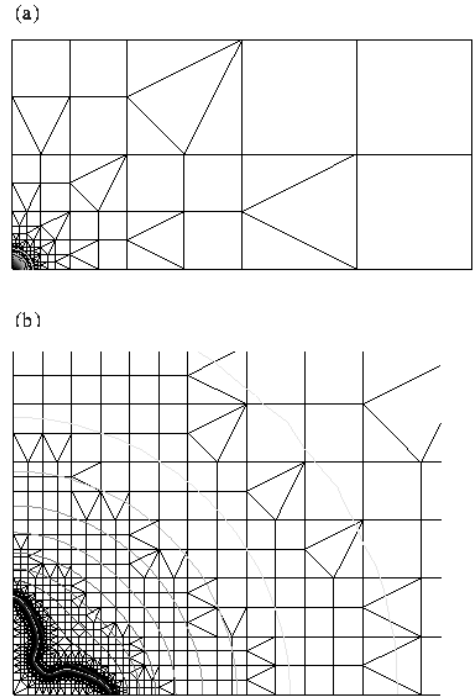}}
\end{center}
\caption{Dendrite mesh and isotherms 
for undercooling $\Delta T = 0.1$. (a) shows the full
domain whose dimensions are $102,400 \times 51,200$. The growing 
dendrite is in the lower left corner.
(b) a close up displaying the dendrite tips, approximately 
$1,300$ units from the origin, while the
temperature field spreads to about $5,000$ units.
}
\label{delta_0.1_pic}
\end{figure}

This is also clearly seen for a dendrite growing at $\Delta=0.1$ in
Fig.~(\ref{delta_0.1_pic}), where the dendrite shape and its
associated field are shown for $\Delta=0.1$ ($D=13$,
$d_0=0.043$, $\epsilon=0.05$, $\Delta x = 0.78$, $dt = 0.08$).
The dendrite arms never became free of each other in this
simulation, causing the observed deviation from solvability
theory shown in Fig.~(\ref{low_delta_vel}).  We note that to  
avoid having the thermal field feel the effect of the sides of the box we perform
our simulations in computational domains for which $L_x \sim (5-10)D/V_n$. To
meet this criterion the simulation for $\Delta=0.1$ was
performed in a $102400 \times 51200$ domain, which is about
$10D/V_n$. 
We note that the ratio of the largest to smallest element size in
this simulation is $2^{17}$.  A fixed mesh having the same
resolution everywhere would contain $9 \times 10^9$ grid points.

We can estimate the time $t^\star$ when the growth of the
dendrite tip crosses over from the transient regime where the
branches interact to that where they become independent by
equating the length of the full diffusion field,
$3(Dt^\star)^{1/2}$, to the length of a dendrite arm,
$V_nt^\star$. This gives the crossover time as
\begin{equation}
t^\star=9D/V_n^2.
\label{cross_over_time}
\end{equation}
The values for $t^\star$ corresponding to
the cases $\Delta=0.45, 0.55 0.65$, and $\Delta=0.25$ and $0.10$  
in Figs.~(\ref{conv_vel0.55}, \ref{sys_vel} and \ref{low_delta_vel}
confirm this scaling. 

These results at low undercooling have important implications
when comparing theory to experimental observations.  In particular, 
since the transient time $t^\star \rightarrow \infty$ as $\Delta \rightarrow 0$, 
it does not appear likely that independent predictions for tip speed 
and radius, as given by solvability,  are likely to be observed experimentally.  
In this regime, the appropriate theory to use to obtain predictions of the 
tip speed and velocity is one which explicitly
takes into account the long range effects of interacting thermal
fields of other branches. Almgren, et al. present one such approach \cite{Alm96}.   
In particular, study of
real dendrites with sidebranches, growing at low undercooling,
will require such treatment.  
In closing, we note that while the independent predictions of tip speed and radius 
deviate from that of solvability theory at low undercoolings, the 
dimensionless {\it stability parameter} $\sigma^*=2d_o D/V_n R^2$  
{\it does} agree within a few percent to solvability theory. 

Further investigation of the tip speeds at low
undercooling, comparison with experiments and new results for 
two-sided directional solidification 
will be reported in forthcoming publications.

\section{Conclusion}

In this paper we present an efficient algorithm used to study
solidification microstructures by adaptive refinement on
a finite element mesh, and solving the phase-field model given
by Eqs.~(\ref{phase-field}).   Our algorithm was made
particularly robust by using dynamic data structures and pointer
variables  to represent our evolving grid. As well, the modular
nature of our code offers an efficient method of expanding the
code to different situations.

We found that our solution time scales with the arclength of the
interface being simulated, allowing simulation of much larger
systems and at very low undercoolings. In particular,
simulations for undercoolings as  low as $\Delta =0.1$ are
quite straightforward in systems larger than $10D/V_n$.
This undercooling represents the upper limit of 
dendrite growth in experiments \cite{Gli84}.

Dendrite tip-velocities at intermediate to high undercoolings
were found to agree with solvability theory to within a few
percent.  At low $\Delta$, we found that the transient
interaction of thermal fields from perpendicular dendrite
branches modifies the tip-velocity from that given by
solvability theory at times shorter than an estimated crossover
time.  Since this crossover time itself becomes larger as
$\Delta$ decreases, it is likely that transient effects will
play a leading role in determining the tip velocity at low
undercooling.  Furthermore, this suggests that at low
$\Delta$ the tip-velocity in the presence of sidebranching will
be different than that predicted by solvability theory. 

Our algorithm is currently being used to examine directional
solidification in models with unequal diffusivities in the
solid and liquid phases.  These results will be presented in 
upcoming publications.

\section*{Acknowledgments}
We thank Wouter-Jan Rappel for providing the Green's function
steady-state code used to test some of our simulations, and
Alain Karma for generously providing us with his unpublished
results. We also thank Robert Almgren and Alain Karma for
helpful discussions of our results at low undercooling. This
work has been supported by the NASA Microgravity Research
Program, under Grant NAG8-1249. We also acknowledge the support 
of the National Center for Supercomputing Applications (NCSA) 
for the use of its computer resources in producing some of our
data.

\section*{References}

\section*{FIGURES}

\begin{figure}[h]
\begin{center}
\leavevmode\mbox{\epsfxsize=2.4in\epsfbox{aniso.epsi}}
\end{center}
\caption{The equilibrium shape of the interface,
for an input anisotropy $\epsilon=0.04$.  The measured  effective anisotropy
$\epsilon_{\rm}=0.041$.
}
\label{aniso_int}
\end{figure}

\begin{figure}[h]
\begin{center}
\leavevmode\mbox{\epsfxsize=2.4in\epsfbox{rich.epsi}}
\end{center}
\caption{Asymptotic steady state velocity as a function of 
minimum grid spacing $\Delta x_{\rm min}$, the for case 
$\Delta=0.55$, $D=2$, $dt=0.016$.
}
\label{rich}
\end{figure}

\end{document}